\input{aipcheck}

\documentclass[
    ,final            % use final for the camera ready runs
  ]
  {aipproc}

\layoutstyle{6x9}

\begin{document}

\title{Present and Future Experiments with Stored Exotic Nuclei at Relativistic Energies
\footnote{Proc. Int. Conf. "Frontiers in Nuclear Structure,
Astrophysics and Reactions", Kos, Greece, 2005}}


\author{H.~Geissel}{
address={Gesellschaft f\"ur Schwerionenforschung GSI,
Planckstra{\ss}e 1, 64291 Darmstadt, Germany}, %
altaddress={Justus-Liebig Universit\"at Giessen,
Heinrich-Buff-Ring 16, 35392 Giessen, Germany}}%
\author{Yu.A.~Litvinov}{
address={Gesellschaft f\"ur Schwerionenforschung GSI,
Planckstra{\ss}e 1, 64291 Darmstadt, Germany}, %
altaddress={Justus-Liebig Universit\"at Giessen,
Heinrich-Buff-Ring 16, 35392 Giessen, Germany}}%
\author{B.~Pfeiffer}{
address={Johannes Gutenberg-Universit\"at,
55128 Mainz , Germany}}%
\author{F.~Attallah}{
address={Gesellschaft f\"ur Schwerionenforschung GSI,
Planckstra{\ss}e 1, 64291 Darmstadt, Germany}} %
\author{G.~Audi}{
address={CSNSM-IN2P3-CNRS, B\^atiment 108, 91405
Orsay Campus, France}}%
\author{K.~Beckert}{
address={Gesellschaft f\"ur Schwerionenforschung GSI,
Planckstra{\ss}e 1, 64291 Darmstadt, Germany}}%
\author{P.~Beller}{
address={Gesellschaft f\"ur Schwerionenforschung GSI,
Planckstra{\ss}e 1, 64291 Darmstadt, Germany}}%
\author{F.~Bosch}{
address={Gesellschaft f\"ur Schwerionenforschung GSI,
Planckstra{\ss}e 1, 64291 Darmstadt, Germany}}%
\author{D.~Boutin}{
address={Gesellschaft f\"ur Schwerionenforschung GSI,
Planckstra{\ss}e 1, 64291 Darmstadt, Germany}, %
altaddress={Justus-Liebig Universit\"at Giessen,
Heinrich-Buff-Ring 16, 35392 Giessen, Germany}}%
\author{T.J.~B\"urvenich}{
address={Max-Planck-Institut f\"ur
Kernphysik, Saupfercheckweg 1, 69117 Heidelberg, Germany}}%
\author{L.~Chen}{
address={Justus-Liebig Universit\"at Giessen,
Heinrich-Buff-Ring 16, 35392 Giessen, Germany}}%
\author{T.~Faestermann}{
address={Technische Universit\"at
M\"unchen, 85748 Garching, Germany}}%
\author{M.~Falch}{
address={Ludwig-Maximilians Universit{\"a}t
M{\"u}nchen, 85748 Garching, Germany}}%
\author{B.~Franzke}{
address={Gesellschaft f\"ur Schwerionenforschung GSI,
Planckstra{\ss}e 1, 64291 Darmstadt, Germany}}%
\author{M.~Hausmann}{
address={Michigan State University,
MI 48824-1321 East Lansing, U.S.A.}}%
\author{E.~Kaza}{
address={Gesellschaft f\"ur Schwerionenforschung GSI,
Planckstra{\ss}e 1, 64291 Darmstadt, Germany}}%
\author{T.~Kerscher}{
address={Ludwig-Maximilians Universit{\"a}t
M{\"u}nchen, 85748 Garching, Germany}}%
\author{P.~Kienle}{
address={Technische Universit\"at
M\"unchen, 85748 Garching, Germany}}%
\author{O.~Klepper}{
address={Gesellschaft f\"ur Schwerionenforschung GSI,
Planckstra{\ss}e 1, 64291 Darmstadt, Germany}}%
\author{R.~Kn\"obel}{
address={Gesellschaft f\"ur Schwerionenforschung GSI,
Planckstra{\ss}e 1, 64291 Darmstadt, Germany}, %
altaddress={Justus-Liebig Universit\"at Giessen,
Heinrich-Buff-Ring 16, 35392 Giessen, Germany}}%
\author{C.~Kozhuharov}{
address={Gesellschaft f\"ur Schwerionenforschung GSI,
Planckstra{\ss}e 1, 64291 Darmstadt, Germany}}%
\author{K.-L.~Kratz}{
address={Johannes Gutenberg-Universit\"at,
55128 Mainz , Germany}}%
\author{S.A.~Litvinov}{
address={Gesellschaft f\"ur Schwerionenforschung GSI,
Planckstra{\ss}e 1, 64291 Darmstadt, Germany}, %
altaddress={Justus-Liebig Universit\"at Giessen,
Heinrich-Buff-Ring 16, 35392 Giessen, Germany}}%
\author{K.E.G.~L\"obner}{
address={Ludwig-Maximilians Universit{\"a}t
M{\"u}nchen, 85748 Garching, Germany}}%
\author{L.~Maier}{
address={Technische Universit\"at
M\"unchen, 85748 Garching, Germany}}%
\author{M.~Mato{\v s}}{
address={Michigan State University,
MI 48824-1321 East Lansing, U.S.A.}}%
\author{F.~Montes}{
address={Michigan State University,
MI 48824-1321 East Lansing, U.S.A.}}%
\author{G.~M\"unzenberg}{
address={Gesellschaft f\"ur Schwerionenforschung GSI,
Planckstra{\ss}e 1, 64291 Darmstadt, Germany},%
altaddress={Johannes Gutenberg-Universit\"at,
55128 Mainz , Germany}}%
\author{C.~Nociforo}{
address={Gesellschaft f\"ur Schwerionenforschung GSI,
Planckstra{\ss}e 1, 64291 Darmstadt, Germany}}%
\author{F.~Nolden}{
address={Gesellschaft f\"ur Schwerionenforschung GSI,
Planckstra{\ss}e 1, 64291 Darmstadt, Germany}}%
\author{Yu.N.~Novikov}{
address={St. Petersburg Nuclear Physics Institute,
188300 Gatchina, Russia}}%
\author{T.~Ohtsubo}{
address={Niigata University, 950-2181 Niigata, Japan}}%
\author{A.~Ostrowski}{
address={Johannes Gutenberg-Universit\"at,
55128 Mainz , Germany}}%
\author{Z.~Patyk}{
address={Soltan Institute for Nuclear Studies,
00681 Warsaw, Poland}}%
\author{W.~Pla\ss}{
address={Justus-Liebig Universit\"at Giessen,
Heinrich-Buff-Ring 16, 35392 Giessen, Germany}}%
\author{M.~Portillo}{
address={Michigan State University,
MI 48824-1321 East Lansing, U.S.A.}}%
\author{T.~Radon}{
address={Gesellschaft f\"ur Schwerionenforschung GSI,
Planckstra{\ss}e 1, 64291 Darmstadt, Germany}}%
\author{H.~Schatz}{
address={Michigan State University,
MI 48824-1321 East Lansing, U.S.A.}}%
\author{C.~Scheidenberger}{
address={Gesellschaft f\"ur Schwerionenforschung GSI,
Planckstra{\ss}e 1, 64291 Darmstadt, Germany}}%
\author{J.~Stadlmann}{
address={Gesellschaft f\"ur Schwerionenforschung GSI,
Planckstra{\ss}e 1, 64291 Darmstadt, Germany}}%
\author{M.~Steck}{
address={Gesellschaft f\"ur Schwerionenforschung GSI,
Planckstra{\ss}e 1, 64291 Darmstadt, Germany}}%
\author{B.~Sun}{
address={Gesellschaft f\"ur Schwerionenforschung GSI,
Planckstra{\ss}e 1, 64291 Darmstadt, Germany}}%
\author{K.~Takahashi}{
address={Max-Planck-Institut f\"ur
Kernphysik, Saupfercheckweg 1, 69117 Heidelberg, Germany}}%
\author{G.~Vorobjev}{
address={Gesellschaft f\"ur Schwerionenforschung GSI,
Planckstra{\ss}e 1, 64291 Darmstadt, Germany},%
altaddress={St. Petersburg Nuclear Physics Institute,
188300 Gatchina, Russia}}%
\author{H.~Weick}{
address={Gesellschaft f\"ur Schwerionenforschung GSI,
Planckstra{\ss}e 1, 64291 Darmstadt, Germany}}%
\author{M.~Winkler}{
address={Gesellschaft f\"ur Schwerionenforschung GSI,
Planckstra{\ss}e 1, 64291 Darmstadt, Germany}}%
\author{H.~Wollnik}{
address={Justus-Liebig Universit\"at Giessen,
Heinrich-Buff-Ring 16, 35392 Giessen, Germany}}%
\author{T.~Yamaguchi}{
address={Saitama University, 338-8570 Saitama, Japan}}%
%
%%%%%%%%%%%%%
%
\begin{abstract}
Recent progress is presented from experiments on masses and
lifetimes of bare and few-electron exotic nuclei at GSI.
Relativistic rare isotopes produced via projectile fragmentation
and fission were separated in flight by the fragment separator FRS
and injected into the storage ring ESR. This worldwide unique
experimental technique gives  access to all fragments with
half-lives down to the microsecond range. The great research
potential is also demonstrated by the discovery of new isotopes
along with simultaneous measurements of mass and lifetime.
Representative results from time-resolved Schottky mass
spectrometry are compared with modern theoretical predictions. The
measured isospin dependence of pairing-gap energies is not
reproduced by  conventional mass models. The first direct
observation of bound-state beta decay has been achieved. Single
particle decay measurements and the continuous recording of both
stored mother and daughter nuclei open up a new era for
spectroscopy. The combination of stochastic and electron cooling
has allowed us to measure with Schottky analysis for the first
time short-lived isomers. The future international NUSTAR facility
at FAIR consisting of a new large-acceptance in-flight separator
(Super-FRS) will be an ideal tool to study the r- and rp-process
nuclei.
\end{abstract}
%
%%%%%%%%%%%
\maketitle
%%%%%%%%%%%
%
%%%%%%%%%%%%%%%%%%%%%%%%%%%%%%%%%%%%%%%%%%%%
%% MAINMATTER
%%%%%%%%%%%%%%%%%%%%%%%%%%%%%%%%%%%%%%%%%%%%

\section{Experimental}

The combination of the in-flight separator FRS \cite{FRS} and the
cooler storage-ring ESR \cite{ESR} at GSI provides unique
experimental conditions with bare and few-electron ions for all
elements up to uranium. Relativistic fragments of several hundreds
MeV/u were produced and separated in flight with the FRS. The
separated reaction products were injected into the ESR with a
selected fixed magnetic rigidity or fixed velocity depending on
the operating mode of the ring. The fragment regions of interest
were selected by varying the incident energy of the projectiles
extracted from the synchrotron SIS.

Due to their stochastic creation process the fragments have an
inevitable velocity spread on the order of one percent which would
aggravate precision measurements in flight. However, a
storage-cooler ring presents ideal tools to overcome this
disadvantageous property. For example, the large phase space is
reduced by electron cooling which enforces all stored ions to the
same mean velocity and reduces the velocity spread to roughly
$3\cdot10^{-7}$ at low intensities. Time resolved Schottky
analysis is ideally suited to determine the revolution frequency
and the intensity of the circulating ions thus providing mass and
lifetime data, respectively \cite{Lit-NP-05,Oht-PRL-05}.

\begin{figure}[h!]
  \includegraphics[width=\textwidth]{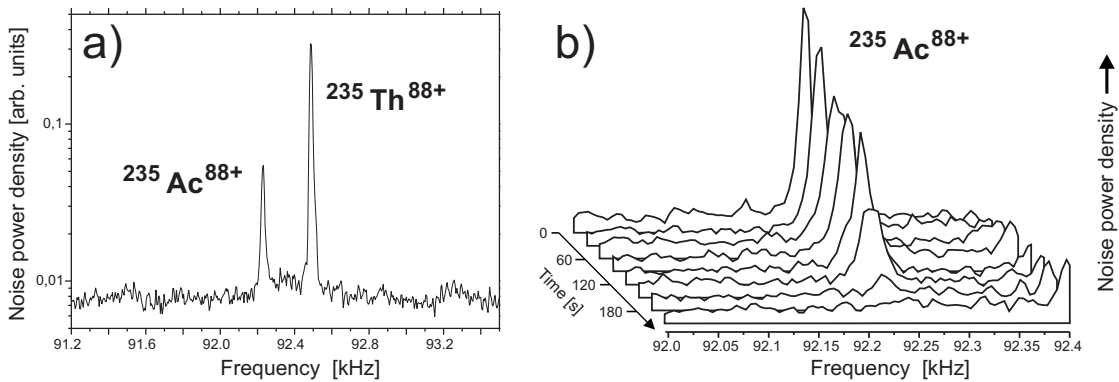}
  \caption{Discovery of the new isotope $^{235}$Ac along with its mass and lifetime measurements
  applying SMS. The mass value has been extracted by calibrating with the well known mass for $^{235}$Th (panel a),
  whereas the half-life has been extracted from the time evolution of the peak area (panel b).}
\label{Figure-235Ac}
\end{figure}

Applying time-resolved Schottky Mass Spectrometry (SMS) we can
correct for possible drifts and thus to achieve higher resolution
plus the feature to record the dynamics of the circulating ions.
This method enables us to achieve an improved mass accuracy of
30~keV and a resolution of $2\cdot10^{6}$
\cite{Lit-NP-05,HG-YL-05}. Tracing in time the isotope peaks in
the Schottky spectra down to a single stored ion, ground or
isomeric states can be assigned even for very small excitation
energies which cannot be resolved under the condition when both
states are simultaneously present \cite{Lit-NP-05}. 285 new and
more than 300 improved mass values of neutron-deficient isotopes
in the range 36$\leq$Z$\leq$85 have been contributed by SMS to the
present knowledge of the mass surface. The research potential of
SMS has been extended in our recent experiments with neutron-rich
uranium fragments where we measured the mass of hitherto unknown
isotopes. This experimental success, illustrated in figure
\ref{Figure-235Ac}, is based on our high sensitivity, resolution,
and unambiguous particle identification.

\section{Results}

\begin{figure}[b!]
  \includegraphics[width=\textwidth]{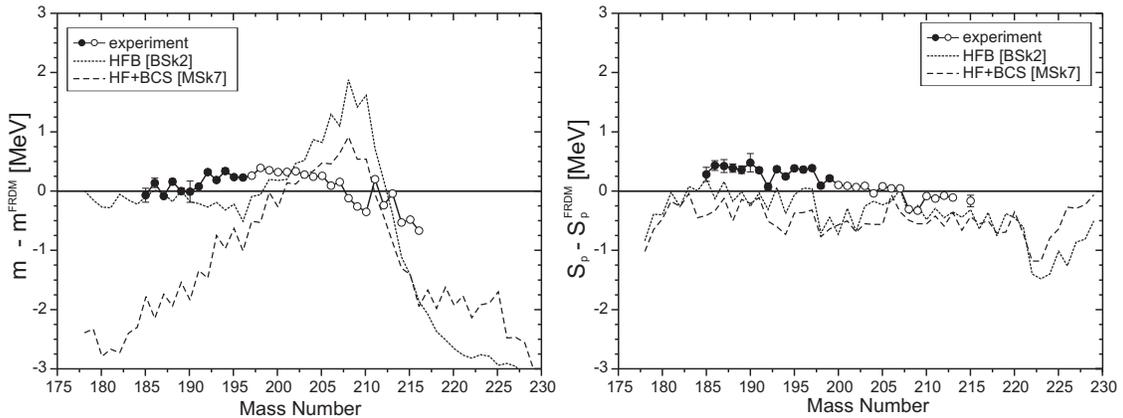}
  \caption{Mass values
(left panel) and proton separation energies (right panel)  of Bi
isotopes compared with the HFB \cite{Go-PRC66},
HF+BCS\cite{Go-PRC66} and FRDM\cite{Mo-ADNDT59} mass models. The
full circles correspond to our new mass values \cite{Lit-NP-05}.}
\label{Figure-Sp}
\end{figure}

The large set of new mass values offers seminal comparisons with
theoretical predictions. For example, in figure \ref{Figure-Sp},
the absolute mass values and the one-proton separation energy
($S_p$)  of Bi isotopes are compared to different modern mass
models. An extensive comparison of all our new masses  yields for
the $\sigma_{rms}$ deviations of the S$_p$ values 379, 525, and
280~keV corresponding to the models of HFB\cite{Go-PRC66},
HF+BCS\cite{Go-PRC66} and FRDM\cite{Mo-ADNDT59}, respectively. The
accuracy for the separation energies is almost a factor of 2
better than the prediction for the corresponding absolute masses
which yield $\sigma_{rms}$ values of 650, 960, and 372~keV for the
HFB\cite{Go-PRC66}, HF+BCS\cite{Go-PRC66} and
FRDM\cite{Mo-ADNDT59} models, respectively. Besides this global
characterization in terms of rms deviations the corresponding
comparison for single isotopes over the mass surface covered by
our experiments has been presented in reference \cite{GLA-NP-04}.
\begin{figure}[h!]
\includegraphics[width=\textwidth]{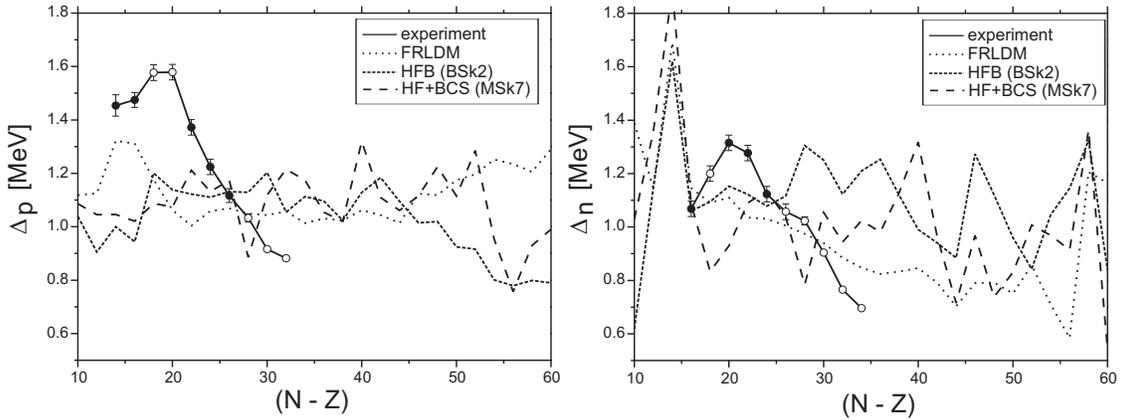}
  \caption{Measured isospin dependence of proton (left panel) and neutron
(right panel) pairing-gap energies for even-even erbium isotopes
compared with nuclear models~\cite{Go-PRC66,Mo-ADNDT59}. The full
circles represent our new mass values \cite{Lit-NP-05}.}
\label{Figure-pairing}
\end{figure}

The accuracy achieved in the present experiment is the basis for
the observation of a new isospin dependence of nuclear pairing
energies. The pairing energies are calculated from measured mass
values using the 5-point formula. In figure \ref{Figure-pairing}
the experimental pairing results are compared to the model
predictions and it is clearly seen that the observed isospin
dependence cannot be reproduced by the current mass models. This
conclusion holds also for the other elements covered in our
measured range of masses \cite{YL-NN03}.

Exotic nuclei with half-lives shorter than the cooling time can be
investigated with time-of-flight techniques operating the ESR  in
the isochronous mode. For isochronous mass spectrometry (IMS)
\cite{HAB00} a special ion-optical setting causes that the
revolution frequency of an ion species is independent of its
velocity spread. Previously, IMS was successfully applied in
smaller isotope regions with many reference masses in the
corresponding spectra \cite{HAB00,SHA-PL-04}. In these pilot
experiments the mass resolution achieved was $1.1\cdot10^{5}$
(FWHM) and the accuracy about 100-500 keV. Recently, a large
number of new neutron-rich masses of fission fragments has been
measured \cite{Matos-PhD}. The analysis of the data is still in
progress and represents a special challenge due to missing
reliable reference masses in this area and secondly due to the
fact that the isochronous condition is strictly fulfilled only for
a part of the stored fragments. An example of IMS results with
nuclear astrophysical relevance is illustrated in figure 4. The
new mass values contribute to the astrophysical calculations of
the rp-process \cite{Schatz-PRL2001}. The possible contribution of
exact neutron separation energies to nuclear astrophysics is
illustrated in figure 5.

\begin{figure}[b!]
\begin{minipage}[b!]{80mm}
\includegraphics*[width=7.cm]{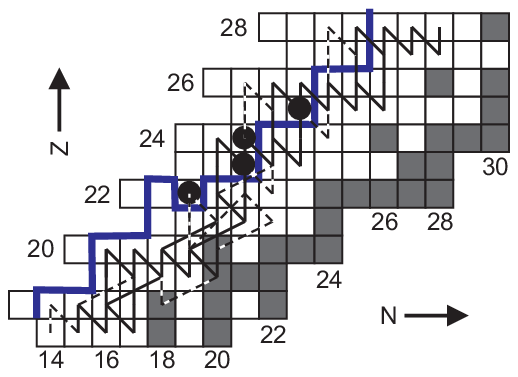}
\end{minipage}
%----------------
\hspace{\fill}
%----------------
\begin{minipage}[b!]{65mm}
{\footnotesize {\bf FIGURE 4.}~~Short-lived masses measured  along
the rp-process \cite{SHA-PL-04} contribute to improve model
calculations \cite{Schatz-PRL2001}.} \label{Figure-rp}
\addtocounter{figure}{1}
\end{minipage}
\end{figure}
\begin{figure}[t!]
\begin{minipage}[t!]{80mm}
\includegraphics*[width=7.0cm]{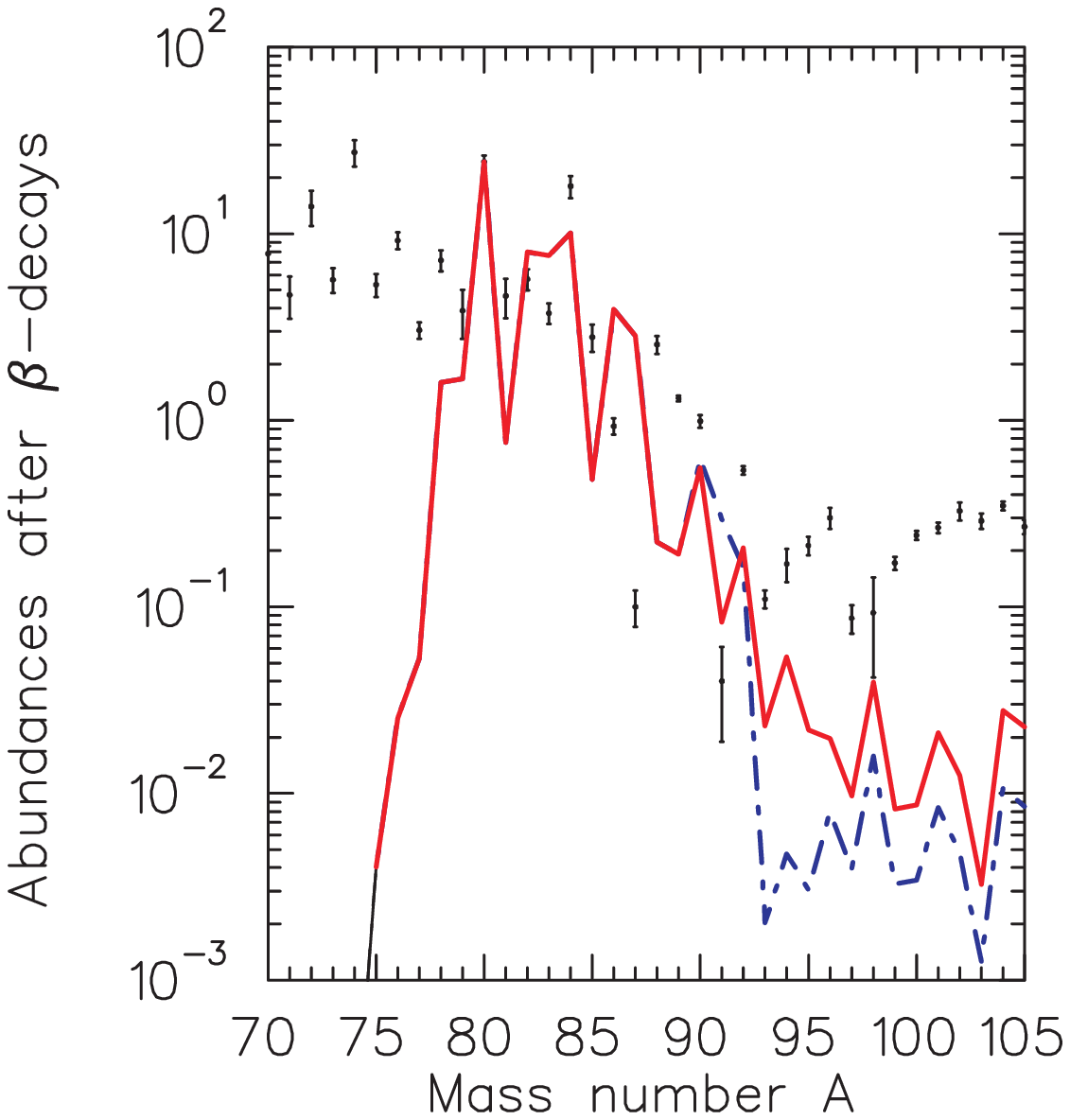}
\end{minipage}
%----------------
\hspace{\fill}
%----------------
\begin{minipage}[t!]{65mm}
{\footnotesize {\bf FIGURE 5.}~~Illustration of the influence of a
single S$_n$ value on the abundance of r-process nuclei. In this
case only the S$_n$ value of $^{93}$Br was varied by 1 MeV (see
two curves) in the HFB \cite{Go-PRC66} calculation. The assumed
conditions for the classical r-process were  10$^{20}$
neutrons~/~cm$^{3}$ and a temperature T$_9$=1.35 K. The measured
Solar System abundances are shown by full circles. }
\label{abundance} \addtocounter{figure}{1}
\end{minipage}
\end{figure}
\begin{figure}[b!]
  \includegraphics[width=14.cm]{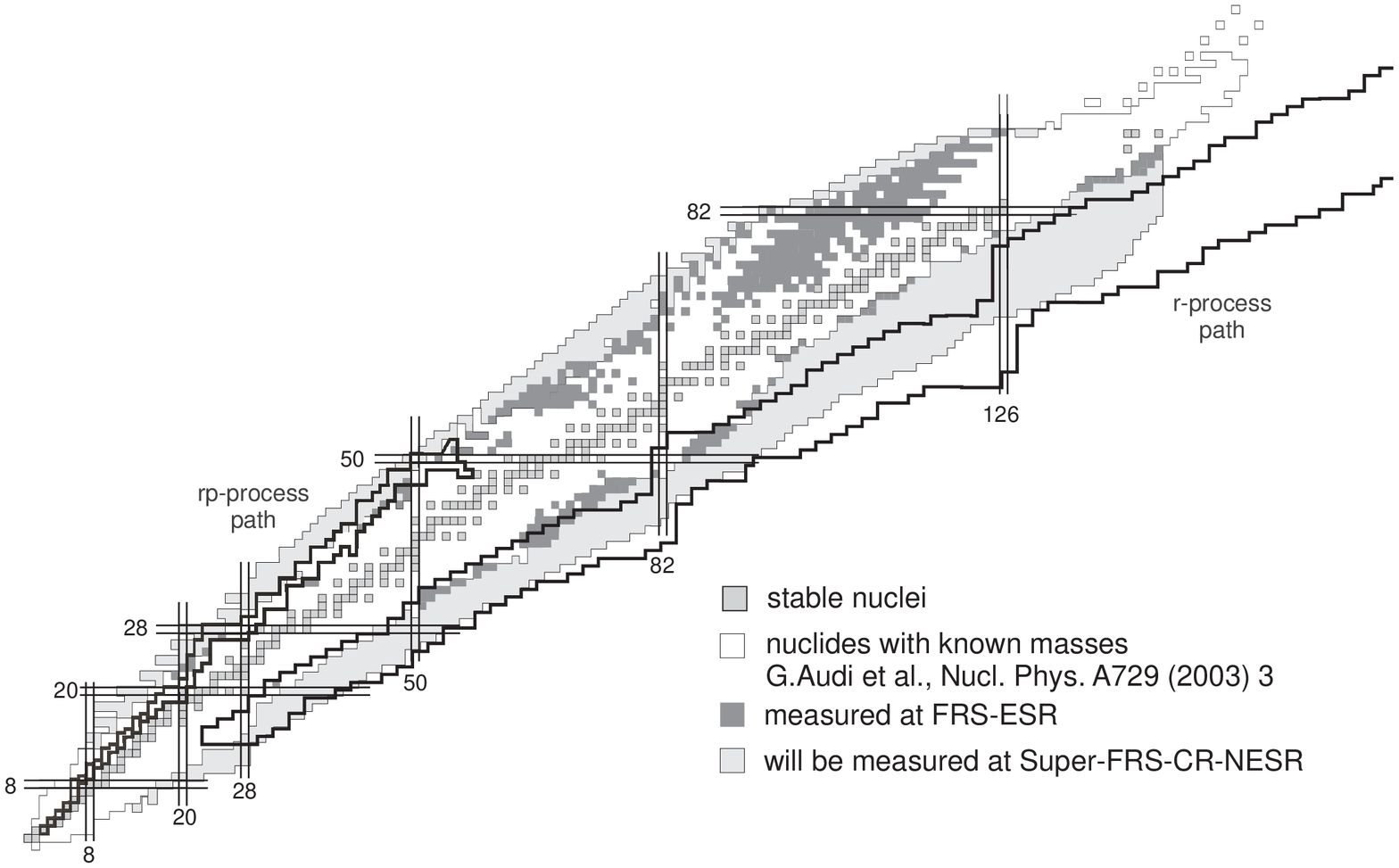}
  \caption{Masses presently covered by FRS-ESR experiments and future range with
  FAIR \cite{CDR}. For the future the following limits have been assumed: 10$^{12}$ projectiles~/~s,
   1 fragment~/~d, T$_{1/2}$ $\geq$10$^{-6} s$.
     The classical r-process corridor is included for orientation. Different neutron
  densities (10$^{20,23,26}$ n~/~cm$^3$) and a temperature T$_9$=1.35 have been assumed.
  The rp-path has been included according to reference \cite{Schatz-PRL2001}.}
  \label{Figure-ILIMA}
\end{figure}
Stored exotic nuclei circulating in the ESR offer unique
perspectives for decay spectroscopy \cite{IGN95,LAB03}. The
half-life of the stored nuclei can be measured by detecting the
mother and daughter nuclides using the difference of their
magnetic rigidity (B$\rho$).  If the B$\rho$ difference is less
than 2.5 \% both nuclei orbit in the storage ring and can be
observed in the same Schottky spectrum. For larger B$\rho$
differences the daughter species leave the closed orbit and can be
detected in a dispersive magnetic dipole stage of the ESR lattice.
The possibility to investigate bare nuclei allows the measurement
of decay properties under the ionization conditions of hot stellar
plasmas. For the first time bound and continuum $\beta^-$ decay
have been simultaneously measured in the laboratory
\cite{Oht-PRL-05}.  The combination of stochastic pre-cooling
\cite{NBC00} and electron cooling yields access to the
spectroscopy of hot fragments with lifetimes down to circa one
second. For example, we have measured the ground and isomeric
states of $^{207}$Tl$^{81+}$ and  the bound-state beta daughter
$^{207}$Pb$^{81+}$. The half-life of the $^{207m}$Tl$^{81+}$
isomer was determined from the evolution in time of the area of
the corresponding peak in the Schottky spectrum. The determined
experimental value for bare $^{207m}$Tl fragments in the rest
frame is 1.47$\pm$0.32~s which is in excellent agreement with the
calculated prolongation (1.52 s) due to the complete suppression
of the internal conversion decay branch \cite{Boutin-PhD-05}.

\section{Summary and Outlook}

In summary, we have demonstrated that the experiments with stored
exotic nuclei at relativistic energies have opened up a new era
for mass and lifetime measurements. Our accurate mass measurements
with cooled nuclei contribute to improve the  theoretical models.
The unique experimental condition to select the atomic charge
states of the fragments down to bare ions can yield new
perspectives for beta decay spectroscopy in general and
specifically for bound-state beta decay. The conditions to provide
pure isomeric beams have been successfully established in our
recent $^{140}$Pr run \cite{Frontiers-05}.

Presently, we have severe intensity limitations for the primary
beams and also in the injection efficiency  of the ESR for hot
fragments. The future international Facility for Antiprotons and
Ion Research FAIR \cite{CDR} will overcome these shortcomings. A
new double-ring synchrotron system (100~/~300 Tm) will accelerate
ions up to uranium with intensities of 10$^{12}$. The beam of
stable isotopes will be converted to rare isotopes with a
large-acceptance superconducting fragment separator (Super-FRS)
\cite{Super-FRS} which will efficiently handle also the large
phase space of the fission fragments. A dedicated storage ring
system will collect, store and cool the fragment beams with minor
losses. These new facilities will allow us to substantially extend
the nuclear physics research and also the major set of
astrophysical relevant nuclei in the nucleosynthesis paths can be
studied for the first time, see figure \ref{Figure-ILIMA}.

\end{document}